# Cost objective PLM and CE

N. Perry and A. Bernard
*Institut de Recherche en Communications et Cybernétique de Nantes (IRCCyN)*
*UMR CNRS 6597, BP 92101, 44321 Nantes Cedex 3, France*

ABSTRACT: Concurrent engineering taking into account product life-cycle factors seems to be one of the industrial challenges of the next years. Cost estimation and management are two main strategic tasks that imply the possibility of managing costs at the earliest stages of product development. This is why it is indispensable to let people from economics and from industrial engineering collaborates in order to find the best solution for enterprise progress for economical factors mastering.
The objective of this paper is to present who we try to adapt costing methods in a PLM and CE point of view to the new industrial context and configuration in order to give pertinent decision aid for product and process choices. A very important factor is related to cost management problems when developing new products. A case study is introduced that presents how product development actors have referenced elements to product life-cycle costs and impacts, how they have an idea bout economical indicators when taking decisions during the progression of the project of product development.

## 1 INTRODUCTION

Concurrent engineering taking into account product life-cycle factors seems to be one of the industrial challenges of the next years. Companies have to elaborate more accurate an reactive plans in order to be able to be in accordance with economical challenges.

Cost estimation and management are two main strategic tasks that imply the possibility of managing costs at the earliest stages of product development.

This is why it is indispensable to let people from economics and from industrial engineering collaborates in order to find the best solution for enterprise progress for economical factors mastering.

But such target is not so easy to define and to characterize. So, consequently, this is always difficult to understand the different mechanisms that are on the critical path on each of the enterprise processes.

The objective is not to re-invent economical evaluation methods but to try to adapt them to the new industrial context and configuration in order to give pertinent decision aid for product and process choices.

A second very important factor is related to cost management problems when developing new products. How to let product development actors have referenced elements to product life-cycle costs and impacts? How to let them have idea bout economical indicators when taking decisions during the progression of the project of product development?

The following paragraphs give some fundamentals about economic methods related to the determination of value chains and the evaluation of industrial costs. An industrial application to casting industry is described and argues as a first application of our approach for economical factors evaluation and control.

## 2 CONTEXT EVOLUTION

The new international market context increases, for the design and manufacturing enterprises, the need of reactivity and agility. However, customer's requirements change: high quality, shorter time to market, innovative and custom products, various services and of course lower prices. For a few decades, the companies have faced significant environment modifications and then had to adapt their management tools, mainly cost management tools.

First, the concept of enterprise evolves to a more dynamic structure that implies the necessity of new methods for performance evaluation and mainly cost estimation.

The development of automation in the production's operations involves the reduction of the labour direct costs. In opposition, the indirect costs, related for example to the maintenance of the reconfigurable manufacturing systems, are more important.

However, in the traditional accountancy systems (method based on homogeneous sections), the overheads are often arbitrary. This does not have any incidence when the percentage of overheads is weak (the impact on the unit costs can be negligible); for example, the time cost of labour can be raised by a coefficient representing a quota of the administration expenses. But when these indirect parts, instead of being residual, represent the major part of the costs, it is necessary to apply finer methods of analysis.

Moreover, the organization of the companies changed. Due to the just in time methodologies, the old functional structure, which means a vertical hierarchical management, changed in a horizontal management including a higher coordination between the services. Therefore, the functional cost division is no longer useful for management. For example, the cost of an after-sale, use to be linked with the commercial function, can be caused by a manufacturing defect; consequently these are the process but not the products that had to be reconsidered. The activities that related to the concept of "value chain", popularised by Mr. Porter (Porter & al. 1995, Ambec & Barla 2002), made much in the awakening of these phenomena.

Second, a widen competing field has to be taken into consideration.

The growing width of the competing field led the companies to reduce on the one hand the times of decision (through the availability of correct economic information at the good time.), and on the other hand the risks of bad decisions (complete visibility of the impacts during the totality of the product life cycle). Therefore, it is essential to have reliable metrics making it possible to quantify the costs and times a priori resulting from an evolution from the products/process at one unspecified time of their life-cycle. Such indicators, which integrate the technique and economic points of view with various steps of the product/process life cycle do not yet exist, or if they do, give an incomplete or not reliable answer. In 1998, this report was a prelude to the formulation of the project METACOG (Methodology of design to total cost objective, part of the French national PROSPER project), which aims to define these metrics.

Third, a horizon evolution has also implied many consequences in the methods and factors for cost evaluation.

Whereas since its beginnings, the calculation of costs is used to improve knowledge of the products costs in production, its application gradually moved towards the future products cost. This change of object does not come out from developments in the technology of costs calculation but from a change in the comprehension and modelling of costs. Then the consideration of the life cycle targets and factors seems an extension, on a longer horizon, of the movement started in the seventies on the costs management tools. Their construction is contingent, taking into account of the economic, the structure and the operating environment of the company.

## 3 COST EVALUATION

First of all, it is essential to clearly define the basic notions of cost management and their evolutions.
The old way to consider the selling price was calculated according to the cost price plus the margin. But today, the selling price is market driven, depending on the fee, which the customer is ready to pay according to the functions, and the services expected. Consequently, the margin is not used to calculate the selling price but it results from it, equal to the selling price minus the cost price. Consequently, the selling value and the margin must be the objectives, and then an "objective maximum cost price", known as "objective cost" can be evaluated as an objective-selling price minus the objective margin. This is this concept which will drive design services during their study phase.

But, it still remains some problems depending on the concept of a product complete cost price and one the margin applied to find the selling price. So cost management methods tried to deal with all this problems (Brodier 1990, Perrin 1996). For example, the respect of the product cost is the base of the target costing methods' and Activity Based Costing (you have to aim an objective target cost and then make correspond the final product cost after what you can measure the gap between the effective and the objective cost). The respect of the budgets, mainly the studies' budgets, is the second key factor for success at the financial level.

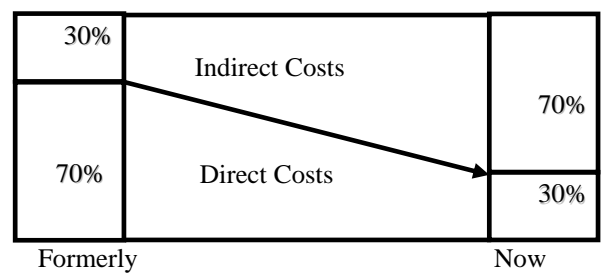

Figure 1: Direct and indirect costs evolution

Formerly, the costs used to take into account a majority of direct costs (often about 70%), i.e. easily affected with the products. The costs not directly easily affected with the products could be the subject of total distributions (the choice of the scale little influenced the result). But now, the proportions between direct to indirect costs are reversed (see figure 1).

In addition, the early phases of the product design decisions' impose more than 85% of the final cost, witch means that the early mistakes or miss evaluations affect in a very large way the costs (see figure 2). The curve above shows the evolution of the expenses really engaged by the company. The main effort for cost management should be focus on the first step of the product life cycle. At the end of the studies, approximately 85% of the complete cost of the program is fixed. Statistically, the share of responsibility in the establishment for the costs is 85% at the engineering and design department, 10% at the office "methods industrialization" and approximately 5 % with manufacture itself and the service purchase.

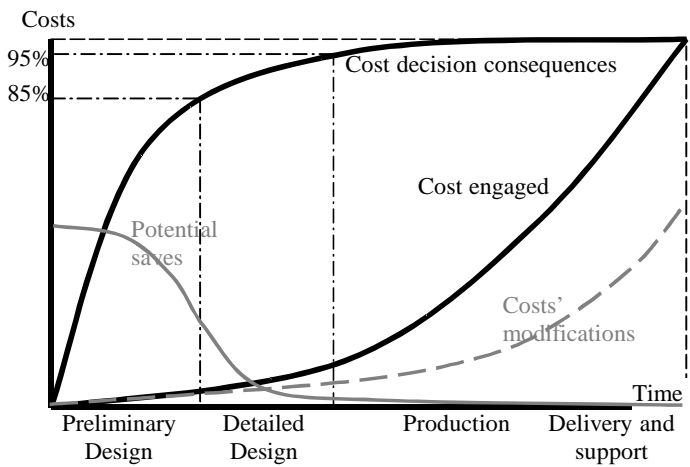

Figure 2: Cost evolution during a project (Alcouffe & Bes 1997)

As indicates it the following figure, more the product's definition advances, less the modifications are easy and the costs increasingly expensive. There is thus an obvious economic interest to concentrate on the search for a product optimisation earlier as possible in the process, the potential saves are significant and the committed costs weak.

Thus, the cost of a product relies on the engineering and design department. Then, the design choices, inexpensive in terms of resources consumption appear in the manufacturing costs when it becomes very hard and expensive to carry out modifications (Bellut 2002, Mevellec 2000).

So, we tried to point out that cost management is hard to handle due to the lack of efficient tools or methods and the very strong effects of decisions in the earlier phases of the product development.

## 4 APPLICATION CASE STUDY

Based on these facts, we developed, with SMC COLOMBIER FONTAINE company, a dynamic method balancing, at the quotation level, the effects of a production on the economical results of the society. This method integrates the enterprise processes in order to manage from the customer requests to the parts to be delivered. The value added unit, chosen for measuring the performance is the time for each step of the product lifecycle in the enterprise. This method follows the product life cycle in the enterprise and is implemented thanks to a collaborative and concurrent engineering approach between all the actors.

SMC COLOMBIER FONTAINE is a French industrial company in the AFE METAL group. The core process is steel sand casting used to manufacture primary parts. They manufacture about 1000 references per year and design an average of 15 new tooling per month.

At the beginning of this work (1999), two factors led the manager develop a complete numerical technology and traceability supporting the product lifecycle in the enterprise (Bernard & al. 2002-a and b):

First the time (then the cost) needed for a new product's industrialisation takes from 20 to 40 hours to complete a CAD study (model the part, the master pattern, the pattern-plates, the cluster and simulate the fill up and the solidification). Furthermore, the same technicians answer to quotations asks, design the tooling, simulate and optimise the complexes parts. Then, the method's service cannot generalize this complete study to all the parts. Furthermore, all these indirect costs took a bigger place in the enterprise financial results due to the important mix of product.

Secondly, the selling price was completely disconnected from the real value added by the enterprise, without taking into account the exact costs of the tooling developments (cores and core boxes, the master patterns...). Finally, the technician globally evaluated the complexity of the part; the price was then calculated on the base of its weight and a ratio depending on the complexity.

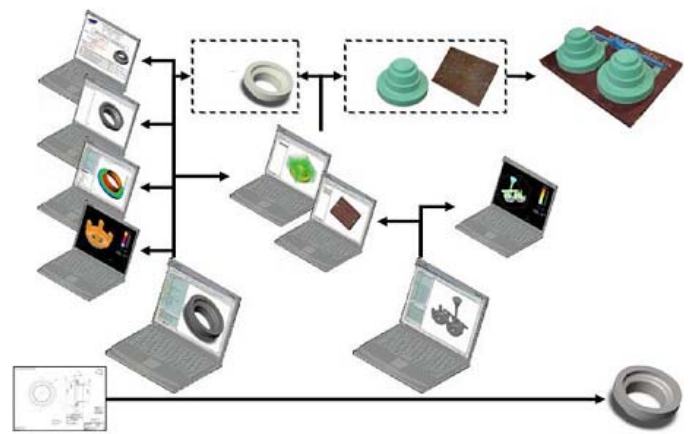

Figure 3: Complete numerical technology (Bernard et al., 2003)

So, in order to reduce the "time to market" (and optimise the efficiency of the tooling office) and to exactly master the costs of the parts, SMC developed a complete numerical technology to support his improvements. This is based on a shared CAD/CAM data and environment, supported by SolidWorks application (see figure 3), and developed for each view required during the PLM in the enterprise (see. figure 4):
- design the part at the quotation level (initial model),
- enriched with the cast rules (skins, machining stock allowance…),
- tooling design (cluster CAD model, master pattern…) and manufacturing (CAM tooling model and NC files),
- process validation and optimisation (fill up and solidification simulations),
- moulds and parts control...

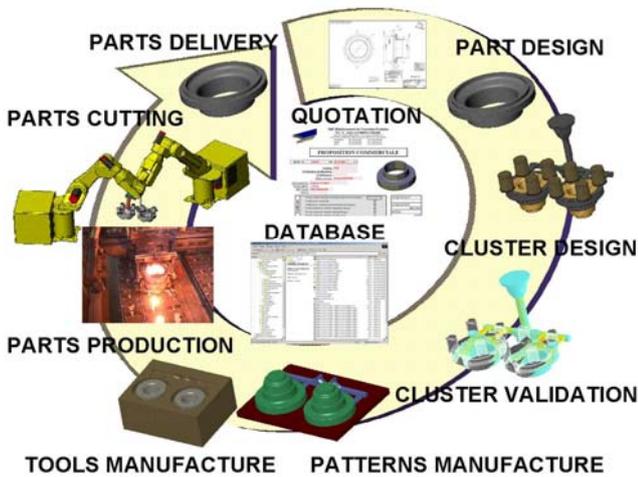

Figure 4: Place of the quotation tool in the numerical loop

This collaborative approach allowed on the one hand a drastic reduction of time to market. Indeed, the old way of development took nearly six weeks before validating the parts prototype whereas a complete integrated solution allow in two weeks and half this validation with the adequate tools (see figure 5) presenting the evolution of the time needed for the several steps of the industrialisation development: study, CAM, wait of pattern maker availability, model machining, master pattern elaboration and prototype validation).

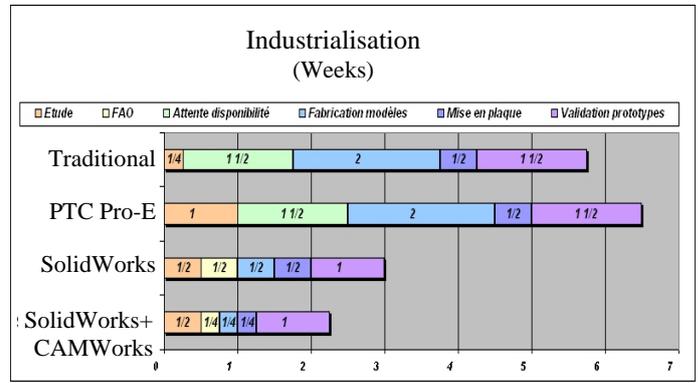

Figure 5: Time to process validation delay

On the other hand, the quotation service can now reply efficiently at each new request with a complete optimised study and a perfect knowledge of the exact time required for the study, the validation and manufacturing of the part. The enterprise processes are completely described in a knowledge database. The time factors, related to the various activities and operations during the product life cycle (time needed for a study, for tooling manufacturing, for the rate of production and so on), are closely looped in order to keep a coherency between the total time evaluation and the exact times spent. This time evaluation is given through a particular Excel's sheet application developed in collaboration with the head director of the enterprise (presented on figure 6) linked with all the processes indicators. This allows estimating the impact on the production capacities and the financial results of the company. This is a very efficient method of capitalizing in detail the characteristic factors of each of the product realization elementary processes.

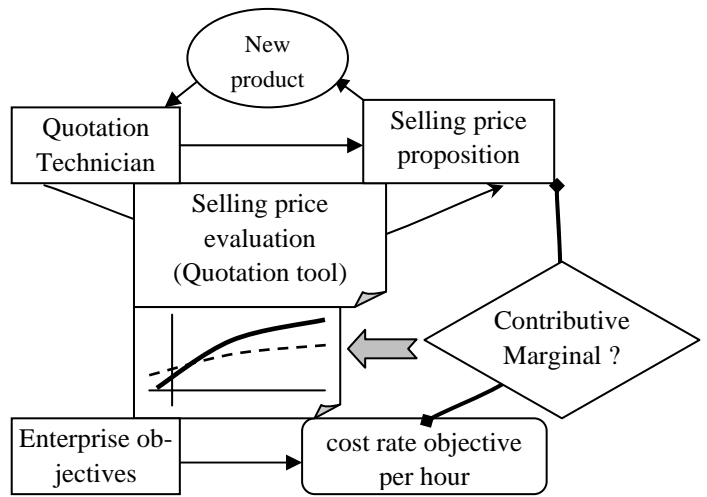

Figure 6: Economical impact estimations

So, when the quotation technician gives a selling price for a new part to be delivered, he sells the time spent for its manufacture with respect to the definition of the characteristic factors of the processes (in addition with the direct costs like raw material for example). It is to be noticed that, with regard to the charge of the process validation (presented above), the time is up to now evaluated and not exactly measured.

In addition, the financial goals of the enterprise, in term of turnover and profit, give a cost rate objective per hour of work, which balances the budget. The technician immediately knows whether, or not, this order will be benefit for the enterprise. Two different kinds of orders can be identified: the "contributive" ones (that cover both fix and variable costs) or the "marginal" ones. It would be easier to only take into consideration the first ones but the production lines should work at their maximum capacities in order to keep a production economic cost. So, some marginal orders are included in the production scheduling to maintain a manufacturing activity for social and economical reasons.

By fulfilling the form, linked with the CAO information's and the manufacturing process (the numerical shared technology starts at this first initial level), the result given by the form is an estimated selling price relying on the time estimated and the objective cost rate. Depending on the final selling price, the order is seen as contributive or marginal. Consequently, each quotation technician had to manage his own impact on the enterprise, taking into account of the effects of his sold parts. The Excel application helps this management with an automatic drawing of curves giving the relative position of all the orders engaged by the technician for the ongoing year and the objective cost rate per hour to adjust the price strategy of the quotations (cf. Figure 7).

Figure 7: Time to process validation delay

## 5 DISCUSSIONS AND CONCLUSIONS

This paper aims to present a cost objective PLM and CE approach and case study in the casting industrial field.

The two main outputs of this research are on one hand the necessity of the definition of effective cost indicators (direct or indirect ones) in order to be able to qualify and quantify the relationships between the process characteristics and their impact on the results of the company. On the other hand, it is of major interest to propose a complete numerical and methodological integration based on a PLM and CE way of thinking. The information system scenarios have been developed in order to be able to deploy a complete closed-loop integrated and collaborative structure and processes.

In parallel, a permanent study of the processes and of the different cost effects allows defining in details the major relationships between the company activity and the value/cost characterization.

Our current research are oriented to the improvement of use and integration of pricing and costing estimating in order to be able to consider economical factors as performance objectives at the same level that all technical and organisational ones. This corresponds to a very present demand from industry mainly due to shorter time-to-market and to the new virtual enterprise structures.

## 6 ACKNOWLEDGEMENTS

The authors would like to thank for their very important contribution to this work SMC Colombiers Fontaine company and more especially MR J.C. Delplace, PhD student, and Mr S. Gabriel, General director of the company and R&D manager of AFE group.